\documentclass[11pt]{article}
\usepackage[margin=1in]{geometry}
\usepackage{amsmath, amsfonts, amssymb}
\usepackage{bm}
\usepackage{graphicx}
\usepackage{parskip}
\usepackage{booktabs}
\usepackage{natbib}
\usepackage{xcolor}
\usepackage[bf,font={small,sl}]{caption}
\usepackage[affil-it]{authblk}

\newcommand{\Z}{Z}

\newcommand{\z}{z}

\title{\textbf{Multiscale modelling of animal movement\\ with persistent dynamics}}
\author{Théo Michelot$^1$\footnote{Email: \texttt{theo.michelot@dal.ca}}, Ephraim M.\ Hanks$^2$}
\affil{$^1$Dalhousie University, $^2$Pennsylvania State University}
\date{}

\begin{document}
\maketitle

\begin{abstract}
    Wild animals are commonly fitted with trackers that record their position through time, and statistical models for tracking data broadly fall into two categories: models focused on small-scale movement decisions, and models for large-scale spatial distributions. Due to this dichotomy, it is challenging to describe mathematically how animals' distributions arise from their short-term movement patterns, and to combine data sets collected at different scales. We propose a multiscale model of animal movement and space use based on the underdamped Langevin process, widely used in statistical physics. The model is convenient to describe animal movement for three reasons: it is specified in continuous time (such that its parameters are not dependent on an arbitrary time scale), its speed and direction are autocorrelated (similarly to real animal trajectories), and it has a closed form stationary distribution that we can view as a model of long-term space use. We use the common form of a resource selection function for the stationary distribution, to model the environmental drivers behind the animal's movement decisions. We further increase flexibility by allowing movement parameters to be time-varying, and find conditions under which the stationary distribution is preserved. 
  We derive an explicit mathematical link to step selection functions, commonly used in wildlife studies, providing new theoretical results about their scale-dependence.
  We formulate the underdamped Langevin model as a state-space model and present a computationally efficient method of inference based on the Kalman filter and a marginal likelihood approach for mixed effect extensions. 
  The approach requires a time discretisation, and we use simulations to investigate performance for various time resolutions of observation. The method works well at fine resolutions, though the estimated stationary distribution tends to be too flat when time intervals between observations are very long. 
\end{abstract}

\section{Introduction}

Wild animals can be tracked using satellite tags, typically resulting in two-dimensional locations $x_1, \dots, x_n$ collected at (possibly irregular) times $t_1 < \dots < t_n$. Various statistical methods have been developed to analyse such data and answer increasingly complex ecological questions about animals' behaviour and their habitat use \citep{hooten2017}. Those methods can broadly be divided into two classes, depending on the spatio-temporal scale of focus: some models describe large-scale patterns of space use by animals, and others focus on small-scale movement patterns. These roughly correspond to the distinction in mathematical biology between Eulerian approaches, which describe the temporal evolution of the spatial distribution of animals, and Lagrangian approaches, which describe the movement dynamics of a single animal \citep{turchin1998}. Although spatial distributions of animals emerge from their small-scale movement decisions, most existing models ignore this mechanism.

A common practical challenge is to jointly estimate movement parameters (e.g., related to speed and directional persistence) and a long-term distribution of space use from animal tracking data. Spatial distributions of animals, often called ``utilisation distributions'', are at the core of wildlife management, and they are used to inform land use regulations and identify important conservation areas. Species distribution models such as resource selection functions are often used for this purpose, but they typically ignore the animals' movement patterns and the resulting autocorrelation, leading to bias and overconfidence \citep{alston2023}. Step selection functions have been proposed as an alternative as they explicitly model animal movement in response to habitat features \citep{forester2009}. However, step selection functions are formulated at the small (Lagrangian) scale, and they do not readily provide a long-term spatial distribution \citep{signer2017, michelot2019linking}. To address this need, multiple analytical and numerical methods have been proposed to compute the utilisation distribution that emerges from a fitted animal movement model \citep{potts2023}, and the most widely-used approach is to simulate many long movement tracks on the landscape and visualise the distribution of sampled locations \citep{signer2024}. However, the absence of an explicit model for the distribution leads to practical and theoretical limitations. In particular, approaches based on step selection functions cannot combine data sets at different scales, their results cannot directly be compared to species distribution models, and the proper quantification of uncertainty in the simulated spatial distributions is challenging and usually ignored.

Recently, it has been proposed that stationary stochastic processes could be used to define movement models that scale up to a known long-term distribution \citep{michelot2019linking}. This is an appealing framework as the stationary distribution models the long-run utilisation of space by the animal, and the stochastic process can be chosen to model the autocorrelation in the animal tracking data. \cite{michelot2020} described a rejection-free Markov chain Monte Carlo algorithm designed to capture basic features of animal movement, but its simple random walk dynamics did not capture directional persistence \citep[present in most animal species;][]{pyke2015}, and it was specified in discrete time such that parameters were tied to the time interval of observation. \cite{michelot2019langevin} proposed using the overdamped Langevin diffusion process as a continuous-time alternative, relaxing dependence on a specific time scale, but that model also failed to capture realistic movement patterns, as the overdamped Langevin diffusion has no directional persistence, its paths are non-differentiable, and its speed is not well-defined (similar to Brownian motion). \cite{whitehead2013} and \cite{wilson_estimating_2018} proposed modelling animal movement as a Markov chain over a discrete spatial grid and computing long-term space use as its stationary distribution, but the movement dynamics and resulting stationary distribution of their model depend on the choice of spatial discretisation. Thus, while those studies provide a proof of concept for the utility of multiscale models that explicitly link movement patterns to a long-term spatial distribution, there is great potential to increase their biological realism and utility.

In this paper, we present a model for animal movement based on the underdamped Langevin diffusion process. The underdamped Langevin process is a highly successful approach in physics to model particle movement \citep{erdmann2000}, and in computational statistics and machine learning to design efficient Markov chain Monte Carlo algorithms \citep{chen2014}, but, to our knowledge, it has not been used in ecology. We show how this model can capture many of the characteristics of fine-scale animal tracking data, has parameters that can be directly interpreted in terms of speed and autocorrelation, and has a known long-run stationary distribution. 
The underdamped Langevin diffusion model gives ecologists a computationally-efficient framework for analysing animal tracking data that allows for familiar inference on movement and resource preferences, with theoretical guarantees about the stationary (utilisation) distribution that are not present in existing step selection models.

\section{Model overview}

We first present a general formulation of the underdamped Langevin process, and then explain how it can be used to model animal movement.

\subsection{The underdamped Langevin process}
\label{sec:model1}

Let $X_t \in \mathbb{R}^d$ be the location and $V_t \in \mathbb{R}^d$ the velocity of an animal at time $t \geq 0$ (usually $d = 2$). The underdamped Langevin process is described by the stochastic differential equations (SDEs),
\begin{equation}
  \label{eqn:model}
  \begin{cases}
    dX_t = V_t\ dt\\
    dV_t = -\gamma V_t\ dt + \sigma^2 \nabla \log [ \pi(X_t) ]\ dt + \sqrt{2 \gamma} \sigma\ dW_t
  \end{cases}
\end{equation}
with initial condition $X_0 = x_0$ and $V_0 = v_0$, where $\pi$ is a log-differentiable function from $\mathbb{R}^d$ to $\mathbb{R}_{>0}$, $\nabla$ is the gradient with respect to location, $(W_t)$ is a $d$-dimensional standard Wiener process, and $\gamma > 0$ and $\sigma > 0$ are parameters governing movement behaviour \citep{cheng2018}. In Equation \ref{eqn:model}, the location is defined as the time integral of the velocity, and the velocity is a mean-reverting process pulled towards higher values of $\pi$ (with some stochasticity). The location process $(X_t)$ therefore combines directional persistence, because velocity is autocorrelated, and attraction to high values of $\pi$. For $d \geq 2$ dimensions, Equation \ref{eqn:model} describes an isotropic process. Under mild regularity conditions, the marginal stationary distribution of the process $(X_t)$ defined in Equation \ref{eqn:model} is proportional to $\pi$ \citep{cheng2018, eberle2019}. In later sections, we will use this property to define a multiscale model where an animal's movement decisions, described by the SDE, are directly linked to the long-run utilisation distribution of the animal.

In physics, Equation \ref{eqn:model} has been used to describe the motion of a particle under an external force measured by $F = \nabla \log \pi$ \citep[][Chapter 10]{risken1996}. In that context, the parameter $\gamma$ is often called the friction coefficient, as larger values imply stronger reversion of the velocity to zero \citep{erdmann2000}. Using the Fokker-Planck approach, the underdamped Langevin process has also been formulated through a partial differential equation for the evolution of the joint probability density function of $X_t$ and $V_t$, known as the Kramers equation \citep{kramers1940, erdmann2000, hadeler2004, gardiner2004}.

Because $(X_t)$ has $\pi$ as its stationary distribution, this process has also been used to build Markov Chain Monte Carlo (MCMC) algorithms to sample from $\pi$ \citep{chen2014}. It is a special case of the general method of \cite{ma2015, ma2019} for specifying SDEs with a given stationary distribution, which they use to define a general family of samplers. Equation \ref{eqn:model} defines an irreversible stochastic process, in contrast with more common reversible processes such as the \emph{overdamped} Langevin diffusion \citep{roberts1996}, and irreversible processes have been shown to explore the target distribution more efficiently in some cases \citep{diaconis2000, bierkens2016}.

\subsection{A multiscale animal movement model}

If the process $(X_t)$ defined in Equation \ref{eqn:model} represents the position of an animal, the function $\pi$ is its long-term distribution over space, analogous to species distribution models widely used in ecology \citep{aarts2012}. By modelling $\pi$ as a function of habitat suitability and other factors that affect the animal's space use (e.g., corridors it needs to travel through), we propose using the underdamped Langevin process as a multiscale model of animal movement, which links the small-scale movement dynamics described by the SDEs (Equation \ref{eqn:model}) to the large-scale spatial distribution $\pi$. \cite{michelot2019langevin} proposed a similar model based on the \emph{overdamped} Langevin diffusion, but the underdamped process presented herein has the advantage of describing realistic small-scale movement dynamics, including directional persistence, differentiable (smooth) paths, and a well-defined speed.

\subsubsection{Movement parameter interpretation}

The parameter $\gamma$ is called the friction coefficient in physics, and larger values lead to more tortuous paths. In animal movement, tortuosity is more likely to be due to an internal state of the animal (e.g., searching for food) than to physical ``friction'' from the environment, but we use this term for convenience. The inverse $1 / \gamma$, sometimes called the relaxation time of the process \citep{risken1996}, is the time interval it takes for the autocorrelation of the velocity to decrease by a factor $e$, such that $3/\gamma$ can be interpreted as the time scale over which the autocorrelation of the velocity decreases by approximately 95\% \citep{johnson2008}. The parameter $\sigma^2$ is most directly interpreted as the variance of the limiting normal distribution of the velocity process (Appendix C). As a result, $\sqrt{\pi/2} \times \sigma$ is the mean speed of movement of the animal in two dimensions \citep{gurarie2011}. Thus, $\gamma$ and $\sigma$ are parameters that control the movement of the animal over short time scales, but leave the stationary utilisation distribution invariant.

\begin{figure}[htbp]
  \centering
  \includegraphics[width=0.5\textwidth]{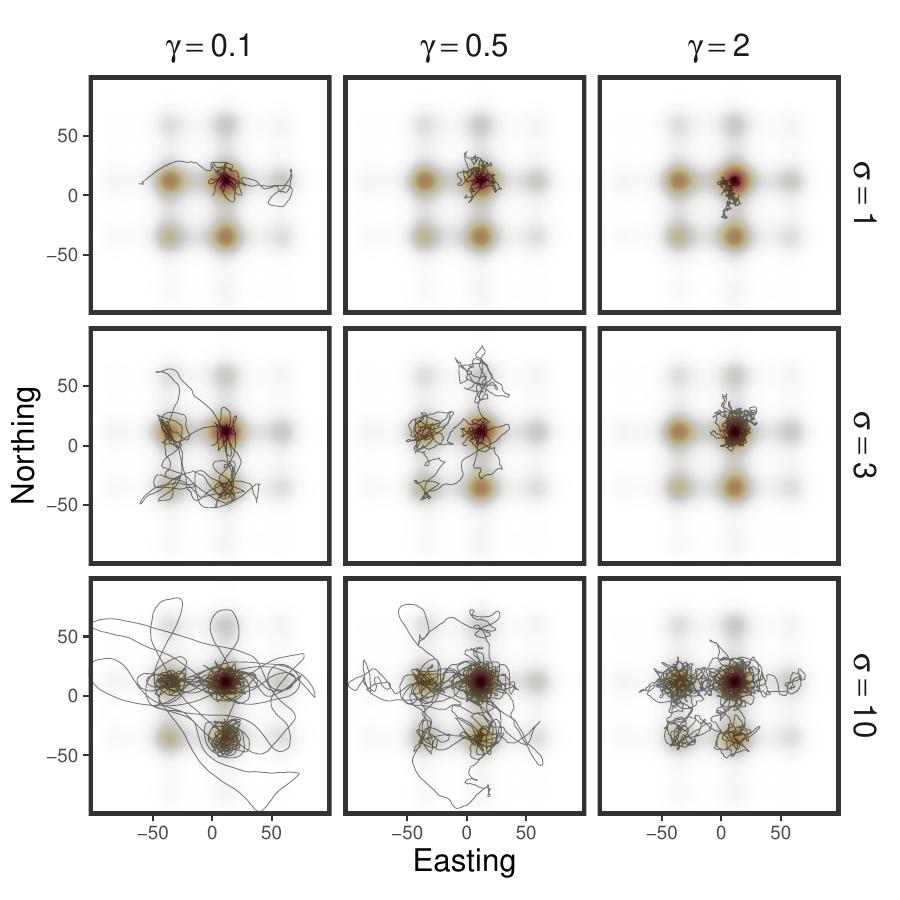}
  \caption{Simulations from the underdamped Langevin process for different values of the movement parameters $\gamma$ (columns) and $\sigma$ (rows). The background colour shows the stationary distribution, which was identical for all simulations.}
  \label{fig:sims}
\end{figure}

Figure \ref{fig:sims} shows simulations from the underdamped Langevin process for different combinations of values for $\gamma$ and $\sigma$, all with the same underlying stationary distribution $\pi$ and over the same time period. It illustrates the flexibility of the model to capture a wide range of movement dynamics, all leading to the same large-scale emergent patterns. In Section \ref{sec:time-var}, we extend this further by allowing $\gamma$ and $\sigma$ to depend on time-varying covariates, e.g., to account for cyclical patterns in an animal's activity, or to be individual-specific.

\subsubsection{Modelling the utilisation distribution}
\label{sec:RSF}

To ensure that it is positive, as is required for a probability distribution, a useful choice for the utilisation distribution $\pi$ is a log-linear model of the form
\begin{equation}
  \label{eqn:RSF}
  \pi(x) \propto \exp\left(\sum_{k=1}^K \beta_k \psi_k(x)\right),\quad x \in \mathbb{R}^d,
\end{equation}
where the $\psi_k$ are known differentiable functions over $d$-dimensional space, and the $\beta_k$ are parameters to be estimated. Due to the linearity of the gradient operator, the gradient term in the SDE for the velocity process can then be written as a linear combination of the gradients of the functions $\psi_k$, i.e., $\nabla \log \pi(X_t) = \sum_{k=1}^K \beta_k \nabla \psi_k (X_t)$. We suggest two possible choices for the functions $\psi_k$, either as spatial covariates that can be used to predict space use (e.g., elevation, vegetation cover), or as basis functions in a semi-parametric approach.

The log-linear form for the utilisation distribution $\pi$ (Equation \ref{eqn:RSF}) is often called a resource selection function when the functions $\psi_k$ are spatial covariates of interest \citep{johnson2006, aarts2012}. The $\beta_k$ coefficients then measure the association between an animal's space use and the measured environmental features, and are often interpreted in terms of strength of ``selection'' or ``avoidance'' \citep{fieberg2021}. This approach can directly be integrated into the underdamped Langevin process: the velocity $V_t$ tends to be ``pulled'' along gradients towards higher values (if $\beta_k > 0$) or lower values (if $\beta_k < 0$) of the covariate $\psi_k$. This provides an intuitive description of an animal's movement decisions, based on the local habitat features, and it is consistent with the ecological theory that animals use resource gradients to navigate their environment \citep{fagan2022}. One important special case is to include $\psi_k(x) = \lVert x-c \rVert^2$ to model home range behaviour around a location $c \in \mathbb{R}^d$, such that (in the absence of other spatial effects) the resulting distribution is Gaussian and centred at $c$. When combined with spatial environmental covariates, this represents an animal that stays centred on the location $c$ but preferentially uses space based on differences in environmental covariates within its home range. 

Alternatively, if spatial covariates are not available, or if they do not fully capture patterns of space use, we can use a semi-parametric model of $\pi$ where the $\psi_k$ are basis functions.  In this context, the weights $\beta_k$ cannot be interpreted separately, and only the resulting function $\pi$ is of interest. This is for example similar to the approaches of \cite{preisler2004}, \cite{russell2018} and \cite{bruckner2020}. In the semi-parametric approach, we may assume that the vector of basis coefficients follows $\beta \sim N(0, \Sigma)$, which can be viewed as a prior distribution or as a smoothness penalty to constrain the shape of the function $\pi$ \citep{wood2017, miller2025}. In the two-dimensional case of animal movement, one could for example use thin plate regression splines, for which both the basis functions $\psi_k$ and the smoothness penalty matrix $\Sigma$ can be obtained using the R package mgcv \citep{wood2003, wood2017}. A model can combine parametric and semi-parametric terms, as is common in spatial modelling where penalised splines are used to capture patterns that cannot be explained by available covariates.

Note that, in practice, environmental covariates are almost always measured and stored over a discrete spatial grid (a ``raster''). The corresponding functions $\psi_k$ are then piecewise constant, and their gradients cannot be used directly in the model (because the slope is zero within each grid cell, and undefined at the boundaries). In this case, we instead consider the bilinear interpolation of the raster, for which the gradient can easily be computed analytically \citep[see e.g.,][]{michelot2019langevin}. Other smoothing or interpolation methods are of course possible, and could be considered in future work. 

\subsection{Time-varying movement dynamics}
\label{sec:time-var}

The dynamics of an animal's movement often change through time, as a result of changes in its behavioural state \citep[e.g., resting, foraging, exploring;][]{morales2004}. As an example, it is common for an animal's behaviour to vary throughout the 24-hour daily cycle. To account for this, two common approaches are (1) to allow for regime switching through a latent Markov chain \citep{blackwell1997, morales2004}, or (2) to specify the movement parameters as functions of time-varying covariates \citep{jonsen2019, michelot2021}. Here, we focus on the latter approach, which is relatively easy to implement, can relate movement patterns to temporal variables, and has been shown in some cases to be a better fit to animal tracking data than Markov-switching models \citep{eisenhauer_flexible_2022}. 

As a general framework for heterogeneous behaviour over time, we propose modelling each movement parameter $\eta_t$ ($\gamma_t$ or $\sigma_t$) to be specified using a combination of fixed and random effects, 
\begin{equation*}
    \log(\eta_t) = a_t^\intercal \alpha + r_t^\intercal u,\quad u \sim N(0, V)
\end{equation*}
where $a_t$ and $r_t$ are vectors of covariates at time $t$, $\alpha$ is a vector of parameters, and $V$ is the covariance matrix of random effects $u$. This formulation allows for linear and non-linear effects of spatiotemporally varying covariates, as well as individual-specific random effects, for example \citep{wand2003, michelot2021}. The matrix $V$ is parameterised in terms of variance parameters $\nu$, which can quantify inter-individual variability (for simple random effects) or control the smoothness of non-linear relationships \citep{wood2017}. In practice, we may have separate coefficients and random effects ($\alpha$, $\nu$, $u$) for $\gamma_t$ and $\sigma_t$.

We follow the approach of \cite{duncan2017} to find conditions on the time-varying parameters $\gamma_t$ and $\sigma_t$ for which the underdamped Langevin diffusion is still stationary. Theorem 1 of \cite{duncan2017} gives a necessary and sufficient condition for a distribution $\rho$ to be a stationary distribution of the (possibly multidimensional) process $Z_t$. Specifically, if the process is the solution to the SDE with drift $a(Z_t)$ and diffusion $\sqrt{2} b(Z_t)$, then proving that $\rho$ is stationary for $(Z_t)$ is equivalent to showing that
\begin{equation}
  \label{eqn:duncan_cond1}
  a(z) = \Sigma(z) \nabla \log \rho(z) + \nabla \cdot \Sigma(z) + \xi(z),
\end{equation}
where $\Sigma(z) = b(z) b(z)^\intercal$, and $\xi$ is a vector field satisfying $\nabla \cdot (\rho \xi)(z) = 0$. In Appendix \ref{app:stationary}, we show that, if $Z_t$ denotes the joint process of location and velocity, Equation \ref{eqn:duncan_cond1} is satisfied for the distribution
\begin{equation}
\label{eqn:rho}
  \rho(z) \propto \pi(x) \exp \left( -\frac{\lVert v \rVert^2}{2\sigma_t^2} \right)
\end{equation}
where $x$ is position and $v$ is velocity, under the conditions that the speed parameter $\sigma_t$ does not depend on the current position or velocity, and the friction parameter $\gamma_t$ does not depend on the current velocity. Equation \ref{eqn:rho} gives the marginal stationary distribution $\pi$ for location, as described previously. If $\sigma$ is constant, then velocity has an isotropic normal stationary distribution with variance $\sigma^2$; otherwise it is a non-stationary process.

This provides a very flexible framework to relate an animal's time-varying movement dynamics to a explicit long-term emerging distribution. Remarkably, dependence of the friction parameter on location does not change stationary behaviour \citep{lim2023}, such that this model can capture effects of spatial (environmental) variables on the degree of movement persistence. It is in principle straightforward to apply a similar method to write the parameters of $\pi$ as functions of time-varying covariates, and this could be used to capture behaviour-dependent habitat preferences \citep[similar to][]{nicosia2017}. However, the time-varying function $\pi$ would not be a stationary distribution for $(X_t)$ any more, and the large-scale distribution of space use may not be available in closed form.

\section{Statistical inference and implementation}
\label{sec:inference}

Animal tracking data consist of observations $x_1, \dots, x_n$ from the location process $(X_t)$, and we want to use those to estimate all model parameters. For simplicity, we first focus on the case where dynamics are not time-varying; then, the vector $\theta$ of parameters to estimate includes the movement parameters $\gamma$ and $\sigma$, and the parameters $\beta_k$ of the stationary distribution $\pi$. We describe inference for the case with time-varying covariates (Section \ref{sec:time-var}) separately in Section \ref{sec:inf-time-var}.

\cite{russell2018} proposed a similar model for the movements of ants, and implemented a Bayesian approach based on the Euler-Maruyama discretisation of the position and velocity processes. \cite{bruckner2020} used an alternative method to derive estimators of the force field and noise in underdamped systems, while correcting for discretisation and observation errors. In this section, we propose another framework of inference based on the Kalman filter. The key challenge is that the velocity process is not observed directly, so we treat it as a latent variable in a state-space model formulation, similar to the approach of \cite{johnson2008} for the integrated Ornstein-Uhlenbeck process. Our new approach is much more computationally efficient than existing Markov chain Monte Carlo approaches.

\subsection{Transition density of discretised process}
\label{sec:transition}

Let $(Z_t) = (X_{t1}, V_{t1}, \dots, X_{td}, V_{td})^\intercal$ be the joint process for location and velocity where $d = 2$ for most wildlife applications. We first consider a realisation from $(\Z_t)$, with values $\z_1, \z_2, \dots, \z_n$ at observation times $t_1 < t_2 < \dots < t_n$, and we define $\Delta_i = t_{i+1} - t_i$ as the $i$-th time interval. There is no general solution to the system of SDEs in Equation \ref{eqn:model} because the dynamics of the process depend on the possibly complex function $\pi$, but an approximate solution can be found based on the assumption that the force term $\nabla \log \pi(X_t)$ is constant over each time interval $[t_i, t_{i+1})$. \cite{cheng2018} derived the transition density of a special case of this discretised underdamped Langevin process (for a particular value of the parameters $\gamma$ and $\sigma$), and we use similar calculations to find the general formulas given below. The derivation is shown in Appendix \ref{app:transition}.

If we denote as $\otimes$ the Kronecker product, $I_d$ the $d \times d$ identity matrix, and $\nabla_j$ the $j$-th component of the gradient with respect to location, then the transition density of the $d$-dimensional isotropic process is
\begin{equation*}
  \Z_{t_{i+1}} \mid \{ \Z_{t_i} = \z_{i} \} \sim N ( \mu_i,\ I_d \otimes Q_i ),
\end{equation*}
where $\mu \in \mathbb{R}^{2d}$ with, for $j \in \{ 1, \dots, d\}$,
\begin{equation*}
    \begin{cases}
        \mu_{i,2(j-1)+1} = x_{ij} + \frac{1-e^{-\gamma \Delta_i}}{\gamma} v_{ij} + \frac{\sigma^2}{\gamma} \left( \Delta_i - \frac{1-e^{-\gamma \Delta_i}}{\gamma} \right) \nabla_j \log \pi(x_i) \\
        \mu_{i,2j} = e^{-\gamma\Delta_i} v_{ij} + \frac{\sigma^2}{\gamma}(1-e^{-\gamma\Delta_i}) \nabla_j\log\pi(x_i)
    \end{cases}
\end{equation*}
and
\begin{equation*}
  Q_i =
  \begin{pmatrix}
    \sigma^2 \left( \frac{2 \Delta_i}{\gamma} - \frac{e^{-2\gamma\Delta_i}}{\gamma^2} - \frac{3}{\gamma^2} + \frac{4e^{-\gamma \Delta_i}}{\gamma^2} \right) & \frac{\sigma^2}{\gamma} \left( 1 - 2 e^{-\gamma \Delta_i} + e^{-2 \gamma \Delta_i} \right) \\
    \frac{\sigma^2}{\gamma} \left( 1 - 2 e^{-\gamma \Delta_i} + e^{-2 \gamma \Delta_i} \right) & \sigma^2 (1 - e^{-2\gamma\Delta_i})
  \end{pmatrix}
\end{equation*}
Over very short time intervals, the slope of $\log \pi$ will, under mild Lipschitz conditions, be almost constant, leading to a good approximation, but the error will be greater at coarser time resolutions. We illustrate implications of discretisation error in simulations in Section \ref{sec:sim}.

\subsection{State-space model formulation}
\label{sec:statespace}

In practice, the velocity $V_t$ is not observed, and we may only observe the location $X_t$, so we take a state-space modelling approach. A state-space model jointly describes the dynamics of an unobserved state process and the mechanism that connects the state to the observed data \citep{durbin2012}. In this context, the underlying state is the joint process $(Z_t)$ for the location and velocity of the animal, and the observation process is just the location $(X_t)$. The model can be written in the form
\begin{center}
  \begin{tabular}{rcl}
    Observation Equation: & & $X_{t_i} = A Z_{t_i}$ \\
    State Equation: & & $Z_{t_{i+1}} = T_i Z_{t_{i}} + B_i h_i + \eta_i$
  \end{tabular}  
\end{center}
where $\eta_i \sim N(0, I_d \otimes Q_i)$ is the process error (with $Q_i$ as defined in Section \ref{sec:transition}), $h_i = \nabla \log \pi(X_{t_i})$, and
\begin{align*}
  & A = I_d \otimes
  \begin{pmatrix}
    1 & 0
  \end{pmatrix},\quad
  T_i = I_d \otimes
  \begin{pmatrix}
    1 & \frac{1 - e^{-\gamma \Delta_i}}{\gamma} \\
    0 & e^{-\gamma \Delta_i}
  \end{pmatrix},\quad \text{and }
  B_i = I_d \otimes
  \begin{pmatrix}
    \frac{\sigma^2}{\gamma} \left( \Delta_i - \frac{1}{\gamma} (1 - e^{-\gamma \Delta_i}) \right) \\
    \frac{\sigma^2}{\gamma} (1 - e^{-\gamma \Delta_i})
  \end{pmatrix}. 
\end{align*}

The observation equation simply represents the fact that velocity is not observed. The state equation is directly derived from the transition density of the discretised underdamped Langevin process given in Section \ref{sec:transition}. Note that, although the state-space formulation could in principle be used to account for measurement error with an additional term in the observation equation, this would not be sufficient to correct for the corresponding errors in $\nabla \log \pi (X_t)$. Therefore, we assume that locations are observed with negligible error, which is standard for telemetry data obtained from the global positioning system (GPS), and we leave the problem of noisy locations open. The analysis of data with substantial measurement error might require a computational algorithm (e.g., Markov chain Monte Carlo) to sample over the true trajectory of the animal \citep[e.g.,][]{brost2015}.

For a state-space model in the form given above, the log-likelihood $\log L(\theta) = \log f(x_1, \dots, x_n \mid \theta)$ can be evaluated using the Kalman filter \citep[][Section 7.2]{durbin2012}. We followed this approach and estimated all parameters using numerical optimisation of the log-likelihood with the R function \texttt{optim} \citep{R2024}. Model parameters can be estimated from multiple independent time series, e.g., movement tracks collected on different animals, by multiplying their likelihoods. 

\subsection{Marginal likelihood for time-varying parameters}
\label{sec:inf-time-var}

If random effect and penalised terms are included in the model for the movement parameters as proposed in Section \ref{sec:time-var}, we propose fitting this model using maximum \emph{marginal} likelihood estimation \citep{skaug2006, wood2017, michelot2021}. Specifically, if we now denote as $\theta$ the fixed effect parameters (which include parameters of the movement, $\alpha$, and of the utilisation distribution, $\beta$), $u$ the random effects, and $\nu$ the variance parameters of random effects, the marginal likelihood is
\begin{equation*}
    L(\theta, \nu) = \int f(x_1, \dots, x_n \mid \theta, u) f(u \mid \nu)\ du
\end{equation*}
where $f(x_1, \dots, x_n \mid \theta, u)$ is given by the Kalman filter, and $f(u \mid \nu)$ is the density function of the multivariate normal distribution of the random effects \citep{skaug2006}. The R package Template Model Builder (TMB) provides a convenient implementation based on the Laplace approximation, leveraging automatic differentiation for efficient optimisation \citep{kristensen2016}. In addition to estimates for $\theta$ and $\nu$, TMB computes best predictors for $u$, following standard random effect modelling approaches \citep[e.g.,][Chapter 7]{fahrmeir_regression_2013}.

\subsection{Uncertainty estimation}

Numerical optimisation of the likelihood function, as proposed above, results in point estimates (or best predictors) and an estimated joint covariance matrix for all fixed parameters and random effects. These can directly be used to obtain standard errors and Wald-type confidence intervals for the fixed and random effect coefficients of the movement parameters ($\alpha$ and $u$ in Section \ref{sec:time-var}), for the variance of random effects, and for the habitat selection parameters of the utilisation distribution ($\beta_k$ in Section \ref{sec:RSF}).

Most often we seek uncertainty estimates for transformed quantities, including for the movement parameters $\gamma$ and $\sigma$ (possibly over a range of covariate values), and for the utilisation distribution $\pi(x)$. We propose a simulation-based approach based on maximum likelihood estimation asymptotics to compute uncertainty estimates:
\begin{enumerate}
    \item Sample $N$ times from the multivariate normal distribution centred on the maximum likelihood estimates, with the joint covariance matrix of fixed and random effects.
    \item For each of the $N$ parameter samples, compute the transformed quantity of interest.
    \item Approximate the standard error as the standard deviation, or confidence intervals using quantiles, based on the $N$ derived quantities.
\end{enumerate}

This approach can be used to quantify the uncertainty on the utilisation distribution $\pi$ over space. In that context, a useful output is the coefficient of variation, given by $\text{SE}(\widehat{\pi}(x))/\widehat{\pi}(x)$ for a location $x$, which quantifies relative error. The utilisation distribution is a key output of habitat selection analyses, which can be used to delineate highly suitable habitats in the context of conservation and management. It is therefore crucial to quantify and communicate its uncertainty in empirical studies. 
By defining a movement model explicitly in terms of the parameters of the utilisation distribution $\pi$, our approach makes it straightforward to produce a coefficient of variation map that explicitly quantifies uncertainty in the estimated distribution, something which is challenging to do with most common movement models.

\section{Links to other movement models}

\subsection{Special and limiting cases}

The underdamped Langevin process includes many widely-used continuous-time movement models as special or limiting cases. The continuous-time correlated random walk (or integrated Ornstein-Uhlenbeck process) of \cite{johnson2008} is obtained when $\nabla \log \pi = 0$, i.e., when the stationary distribution $\pi$ is an improper uniform distribution in space. The model presented here can therefore be viewed as an extension of that model to the case where movement is driven by habitat, similar to the approach of \cite{russell2018}. 

When both $\gamma \to \infty$ and $\sigma \to \infty$ such that $\sigma^2/\gamma$ remains constant, the underdamped Langevin process reduces to the overdamped (first-order) Langevin process and the position process is described by
\begin{equation*}
    dX_t = \zeta^2 \nabla \log[\pi(X_t)]\ dt + \sqrt{2}\zeta\ dW_t
\end{equation*}
where $\zeta = \sigma/\sqrt{\gamma}$ is a diffusion parameter (details in Appendix \ref{app:overdamped}). \cite{michelot2019langevin} used the overdamped process for animal movement, but it has no directional persistence and its paths are non-differentiable (i.e., velocity is ill-defined), making it a less realistic model for fine-scale movement dynamics and for modern movement data collected at fine time resolutions.

\cite{brillinger2011} proposed using potential functions to model animal movement based on an analogy with particle motion. That work briefly presented a deterministic model for position and velocity that is similar to the underdamped Langevin process, but then focused on the overdamped case through the asymptotic argument we presented in the previous paragraph, and most other potential function papers have followed a similar approach \citep{preisler2004}. Some exceptions have considered potential function models for velocity \citep{hanks2011, russell2018, eisenhauer2020}, and those are most similar to the model we propose. While \cite{preisler2004} noted the relationship between the potential function and the stationary distribution, none of these papers made the formal link to resource selection functions and utilisation distributions that we make here.

\subsection{Approximate equivalence to step selection function models}

Step selection functions are one of the most popular models of animal movement and habitat selection \citep{forester2009}. They describe movement in discrete time, and the distribution of the position $X_{i+1}$ at time $t_{i+1}$ given previous positions $X_{1:i}$ is typically in the form $f(x_{i+1} \mid x_{1:i} ) \propto \exp [ \sum_{k=1}^K \eta_k h_k(x_{i+1}, x_{1:i}) ]$, for some environmental or movement variables $h_k$. Typically, $h_k$ can include spatial covariates at $x_{i+1}$ (e.g., elevation), and functions of the distance between $x_i$ to $x_{i+1}$ (to capture speed constraints) and the angle between $x_{i+1} - x_i$ and $x_i - x_{i-1}$ (for directional persistence). A step selection function predicts the distribution of the animal one step ahead, but there are no general theoretical results on the emerging stationary distribution. We provide one avenue towards such results by showing that, for small time intervals, the underdamped Langevin diffusion is approximately equivalent to a step selection function with a specific choice of covariates $h_k$. For simplicity, we denote as $X_1, X_2, \dots$ and $V_1, V_2, \dots$ the values of the processes $(X_t)$ and $(V_t)$, respectively, at regular observation times $t_1 < t_2 < \dots$. We denote as $\Delta = t_{i+1} - t_i$ the constant time interval, and assume that $\Delta$ is small. Details of the full derivation are given in Appendix \ref{app:ssf}.

Combining the transition density for velocity given in Section \ref{sec:transition} with the first-order discretisation $V_i \approx (X_{i}-X_{i-1})/\Delta$, we have the approximation
\begin{equation*}
X_{i+1} \mid x_i, x_{i-1} \sim N \left( 
    x_i + e^{-\gamma \Delta} (x_i - x_{i-1}) + \Delta (1-e^{-\gamma \Delta}) \frac{\sigma^2}{\gamma} \nabla \log \pi(x_i),\ 
    \Delta^2 \sigma^2 (1 - e^{-2\gamma \Delta}) I_d \right) 
\end{equation*}
The density function of the (approximate) conditional distribution of $X_{i+1}$ is therefore
\begin{equation}
\label{eqn:approx_pdf1}
    f(x_{i+1} \mid x_i, x_{i-1}) \propto \exp \left[ -\frac{a^\intercal a}{4\gamma \Delta^3 \sigma^2} \right]
\end{equation}
where $a = (x_{i+1} - x_i) - e^{-\gamma\Delta}(x_i - x_{i-1}) - \Delta (1-e^{-\gamma\Delta}) (\sigma^2/\gamma) \nabla \log \pi(x_i)$, and where the constant of proportionality does not depend on $x_{i+1}$. Expanding, and grouping together the terms that do not depend on $x_{i+1}$ into a constant $C$, we can write
\begin{equation*}
    a^\intercal a = \lVert x_{i+1} - x_i \rVert^2 - 
        2 (x_{i+1} - x_i)^\intercal (x_i - x_{i-1}) -
        2 \Delta (1-e^{-\gamma\Delta}) \frac{\sigma^2}{\gamma} (x_{i+1} - x_i)^\intercal \nabla \log \pi(x_i) +
        C.
\end{equation*}
We denote as $L_i = \lVert x_{i+1} - x_i \rVert$ the Euclidean distance between the two positions $x_i$ and $x_{i+1}$ (usually called step length), and $\varphi_i$ the change in direction between the vectors $x_i - x_{i-1}$ and $x_{i+1} - x_i$ (called the turning angle). By property of the inner product, we have $(x_{i+1} - x_i)^\intercal (x_i - x_{i-1}) = L_{i-1} L_i \cos(\varphi_i)$. Using a linear approximation of $\log \pi$ in a neighbourhood of $x_i$ and $x_{i+1}$ (which is reasonable for small $\Delta$ under standard Lipschitz conditions), we also have $(x_{i+1} - x_i)^\intercal \nabla \log[\pi(x_i)] \approx \log[\pi(x_{i+1})] - \log[\pi(x_{i})]$. Substituting these expressions into $a^\intercal a$, and using the exponential form of $\pi$ proposed in Equation \ref{eqn:RSF}, we get the following approximation from Equation \ref{eqn:approx_pdf1},
\begin{align}
    f(x_{i+1} \mid x_i, x_{i-1}) \propto \exp \Bigg[ 
    \frac{-1}{2\Delta^2\sigma^2(1-e^{-2\gamma\Delta})} L_i^2 + 
    \frac{e^{-\gamma\Delta}}{\Delta^2\sigma^2(1-e^{-2\gamma\Delta})} L_{i-1} L_i \cos(\varphi_i) + \nonumber \\
    \sum_{k=1}^K \frac{(1-e^{-\gamma\Delta}) \beta_j}{\Delta\gamma(1-e^{-2\gamma\Delta})} \psi_k(x_{i+1}) \Bigg]
    \label{eqn:ssf1}
\end{align}

Equation \ref{eqn:ssf1} is in the form of a step selection function that includes covariates specified as (a) the squared step length with a negative coefficient, representing avoidance of long steps due to energetic constraints, (b) the product $L_{i-1} L_i \cos(\varphi_i)$ with a positive coefficient, which captures persistence in movement speed and direction, and (c) the spatial covariates $\psi_1, \dots, \psi_K$, for selection or avoidance of habitat features. This formal link suggests that step selection function models with the specific covariates described above, if observed at very small time intervals, will exhibit short-term dynamics very similar to those of the underdamped Langevin diffusion.

The coefficient in front of $\psi_k(x_{i+1})$ in Equation \ref{eqn:ssf1} is of the order of $\beta_j / (2\gamma\Delta)$, suggesting that the apparent strength of selection estimated by a step selection function might decrease as the time interval of observation $\Delta$ grows, at least over some range of small $\Delta$ values. This is different from the result for step selection functions with no directional persistence, where apparent selection increases with time interval \citep[from $\beta_j/2$ to $\beta_j$;][]{barnett2008, potts2023}. Therefore, we posit that the relationship between time interval and step selection depends crucially on the amount of persistence in an animal's movement, and Equation \ref{eqn:ssf1} provides some insights into its shape for small $\Delta$.

\section{Simulations}
\label{sec:sim}

We assessed the performance of the method of inference presented in Section \ref{sec:inference} using simulations. We defined the true stationary distribution as $\pi(x) = \exp(\beta_1 \psi_1(x) + \beta_2 \psi_2(x) + \beta_3 \lVert x - c \rVert^2)$ where $\psi_1$ and $\psi_2$ were generated at random on a raster grid, to resemble spatial environmental covariates, and where $c$ was the centre of the region. We used the coefficients $(\beta_1, \beta_2, \beta_3) = (2, 5, -10)$, representing positive ``selection'' for the two environmental variables, and a tendency to remain close to the centre (similar to home range behaviour of animals). For the simulation, the movement parameters were set to $\gamma = 1$ and $\sigma = 1$. We simulated 100 trajectories from the discretised underdamped Langevin process on this map, using a very fine time resolution of $\Delta = 0.01$ to avoid any discretisation error in the simulation, between times 0 and 500. Then, we downsampled the data to time intervals of $\Delta \in \{ 0.02, 0.05, 0.1, 0.2, 0.5, 1, 2\}$, and fitted the model on the thinned data.

\begin{figure}[htbp]
  \centering
  \includegraphics[width=\textwidth]{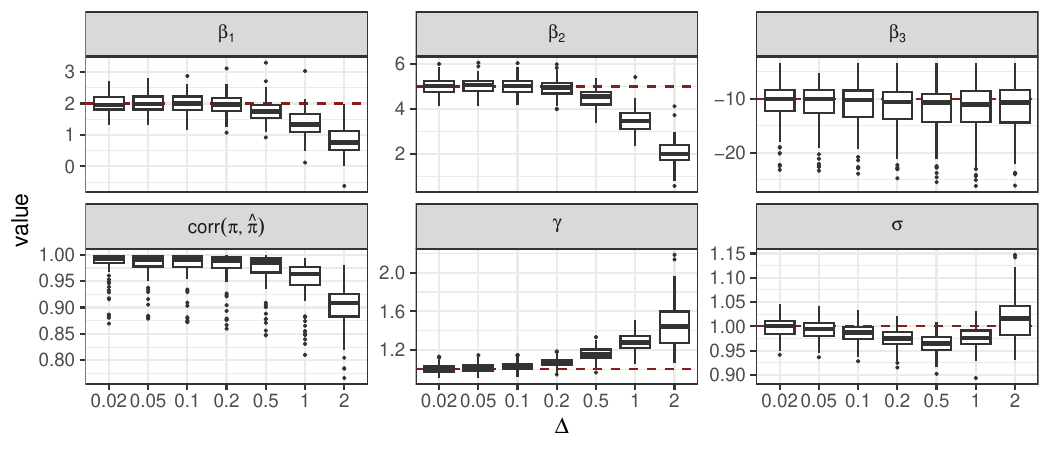}
  \caption{Results of simulation study. Parameter estimates from 100 simulations, for data thinned to different time resolutions $\Delta$. The bottom-left panel shows the correlation between the estimated and true utilisation distributions, measured on a raster grid. The other five panels show the five parameters of the model, with the true values shown as horizontal red lines.}
  \label{fig:sim_res1}
\end{figure}

The results are shown in Figure \ref{fig:sim_res1}. All model parameters were recovered well at short time resolutions. The friction parameter $\gamma$ displayed some moderate positive bias at coarse time resolutions, corresponding to a slight underestimation of the time scale of autocorrelation of the velocity process. The speed parameter $\sigma$ was estimated well even for the longer time intervals. In the model for the stationary distribution $\pi$, the estimates for the coefficients $\beta_1$ and $\beta_2$ (measuring the association between the distribution and two artificial environmental variables) decreased as the time interval of observation increased; for very long intervals, they were estimated close to zero. This is because those parameters are estimated based on the association between the animal's movement and the gradient of the covariates at the start point of each observed step. As the time intervals get longer, the association becomes weaker, and the coefficients are underestimated. Interestingly, the coefficient $\beta_3$ for the squared distance $\lVert x - c \rVert^2$ did not seem to be biased even at coarse resolutions, although the estimation variance increased somewhat. This is probably because $\lVert x - c \rVert^2$ does not change quickly over space, i.e., the assumption that its gradient does not change between two observed locations is approximately correct even for coarser data. As a result, even though $\beta_1$ and $\beta_2$ could not be recovered at long intervals, the overall shape of the utilisation distribution $\pi$ was estimated quite well; the correlation between the true and estimated $\pi$ (measured on a raster grid) was around 0.9 on average even at the coarsest resolution.

From this simulation study, we conclude that the estimation approach is able to recover parameters related to general space use ($\beta_3$), resource selection ($\beta_1$ and $\beta_2$), and movement behaviour ($\gamma$ and $\sigma$) when telemetry data are observed at high temporal resolution. For longer observation intervals, only parameters related to general space use that change slowly over space may be estimated.

\section{Discussion}

The underdamped Langevin model has great potential to obtain a wide range of biological insights from animal tracking data, including estimates of biologically-relevant movement parameters related to speed and persistence (and their relationships to covariates), habitat selection parameters, and spatial distributions with associated uncertainty. The mixed modelling extension allows for a parsimonious formulation to account for inter-individual differences while borrowing information across the population. This is a unifying framework to study the space use of animals across spatiotemporal scales.

It is a common challenge to carry out inference for SDEs from discrete data, and most approaches use approximations such as the Euler-Maruyama discretisation \citep{iacus2008}. Similarly, our approach requires the assumption that the gradient of the stationary distribution $\nabla \log \pi (X_{t_i})$ is approximately constant between two observation times $t_i$ and $t_{i+1}$, and simulations revealed that this can lead to bias for coarse temporal resolutions. We tried replacing $\nabla \log \pi(X_{t_i})$ by $(\nabla \log \pi(X_{t_i}) + \nabla \log \pi(X_{t_{i+1}}))/2$, or by $\nabla \log \pi((X_{t_i} + X_{t_i})/2)$, to account for dependence on the shape of $\pi$ along the whole step (rather than only at the start point), but this did not yield noticeably better results. Various numerical methods have been developed to improve inference for SDEs. A common approach is to use data augmentation (sometimes called ``infill'') to approximate the SDE on a finer time resolution than the observed data to decrease the discretisation error, and marginalise over missing observations using numerical integration \citep{durham2002} or Markov chain Monte Carlo \citep{hanks2017}. In principle, our inferential approach could be extended to this case, with the details of such an implementation being a possible avenue for future work.

The Kramers equation is a partial differential equation describing the evolution of the joint probability density of the position and velocity for the underdamped Langevin process \citep{hadeler2004}. It could be used to study the spatial distribution of an animal over intermediate time steps, theoretically or using simulations \citep[``transient dynamics'';][]{potts2024}. This could for example provide insights into the time scale over which animals explore their environment, which is analogous to the concept of mixing time and convergence speed of stochastic processes. Problems related to exit times and first passage times of the Kramers equation have also been studied to describe how a particle might move between several local minima of the stationary distribution \citep{kramers1940}; this is highly relevant in ecology where animals commonly move in fragmented landscapes where good habitat patches are separated by lower-quality environment.

We only considered the isotropic model where the SDE parameters $\gamma$ and $\sigma$ are scalars, representing the assumption that the spatial dimensions are interchangeable for the animal. We could replace $\gamma$ by a rotation matrix to relax this assumption, similar to the approach of \cite{gurarie2017} and \cite{delporte2024} for the simpler integrated Ornstein-Uhlenbeck process. In particular, \cite{delporte2024} used this approach to capture spatial barriers, such as the effect of land on the movement of marine animals, and this could be integrated into the underdamped Langevin model. Non-isotropic processes could also be explored for animals moving in three dimensions, to represent the higher energetic expenditure typically incurred from vertical movement compared to horizontal movement.

The underdamped Langevin diffusion process has many practical advantages for wildlife ecologists working with movement data. Its continuous-time formulation makes it straightforward to analyse tracking data collected at irregular time intervals, or to compare and combine studies with different sampling frequencies. The extension to time-varying movement parameters accounts for behavioural variations, which are often observed over the course of tracking studies. Most importantly, this multiscale model provides a much-needed mechanistic link between small-scale movement and emergent large-scale distribution, allowing for the estimation of animal distributions from tracking data while accounting for autocorrelation. Our model could directly be used within the joint likelihood framework of \cite{blackwell2024} to combine different types of spatial data (e.g., GPS and camera trap data), which can greatly increase the power of ecological analyses. Other stationary processes within the general class described by \cite{ma2015, ma2019} might provide even better descriptions of animal movement, and we hope that this paper will stimulate more work in this area.

\subsection*{Acknowledgements}
We thank Paul Blackwell for discussions on this topic, and for bringing \cite{ma2019} to our attention.

\bibliographystyle{apalike}
\bibliography{refs}

\vspace{2cm}
\appendix
\begin{center}
  \LARGE\bf Appendices
\end{center}

\renewcommand{\thefigure}{S\arabic{figure}}
\setcounter{figure}{0}

\section{Stationarity}
\label{app:stationary}

We follow the approach of \cite{duncan2017} to find a stationary distribution of the underdamped Langevin diffusion, with a focus on time-varying parameters. Theorem 1 of \cite{duncan2017} gives a necessary and sufficient condition for a distribution $\rho$ to be a stationary distribution of the (possibly multidimensional) process $Z_t$. Specifically, if the process is the solution to the SDE with drift $a(Z_t)$ and diffusion $\sqrt{2} b(Z_t)$, then proving that $\rho$ is stationary is equivalent to showing that
\begin{equation}
  \label{eqn:duncan_cond}
  a(z) = \Sigma(z) \nabla \log \rho(z) + \nabla \cdot \Sigma(z) + \xi(z),
\end{equation}
where $\Sigma(z) = b(z) b(z)^\intercal$, and $\xi$ is a vector field satisfying $\nabla \cdot (\rho \xi)(z) = 0$.

Focusing on a single dimension for convenience, we denote the location as $x$, velocity as $v$, and $z = (x, v)$. Then, the drift and diffusion functions of the underdamped Langevin diffusion with time-varying friction parameter $\gamma_t > 0$ and speed parameter $\sigma_t > 0$ are
\begin{equation*}
  a(z) =
  \begin{pmatrix}
    v \\
    -\gamma_t v + \sigma_t^2 \nabla_x \log \pi(x)
  \end{pmatrix}
  \quad
  \text{and}
  \quad
  b(z) =
  \begin{pmatrix}
    0 \\
    \sqrt{\gamma_t} \sigma_t
  \end{pmatrix}.
\end{equation*}
Note that the factor $\sqrt{2}$ is taken out of the function $b(z)$, following \cite{duncan2017}. In the following, we show that the condition in Equation \ref{eqn:duncan_cond} is satisfied for the distribution
\begin{equation*}
  \rho(z) = K \pi(x) \exp \left( -\frac{v^2}{2\sigma_t^2} \right)
\end{equation*}
where $K$ is a normalising constant, under some constraints on the (possibly time-varying) parameters $\gamma_t$ and $\sigma_t$. Note that we distinguish between the two-dimensional distribution $\rho$ and the one-dimensional marginal distribution $\pi$ for position.

\subsection{Step 1: $\Sigma(z) \nabla \log \rho(z)$}

We have
\begin{equation*}
  \Sigma(z) = b(z) b(z)^\intercal =
  \begin{pmatrix}
    0 & 0 \\
    0 & \gamma_t \sigma_t^2
  \end{pmatrix}
\end{equation*}
and
\begin{equation*}
  \log \rho(z) = \log(K) + \log \pi(x) - \frac{v^2}{2\sigma_t^2}
\end{equation*}
Assuming that $\sigma_t$ is constant with respect to $x$ and $v$,
\begin{equation*}
  \begin{cases}
    \nabla_x \log \rho(z) = \nabla_x \log \pi(x) \\
    \nabla_v \log \rho(z) = - \frac{v}{\sigma_t^2}    
  \end{cases}
\end{equation*}
where $\nabla_x = \partial/\partial x$ and $\nabla_v = \partial / \partial v$. We find
\begin{equation*}
  \Sigma(z) \nabla \log \rho (z) =
  \begin{pmatrix}
    0 & 0 \\
    0 & \gamma_t \sigma_t^2
  \end{pmatrix}
  \begin{pmatrix}
    \nabla_x \log \pi(x) \\
    -v / \sigma_t^2
  \end{pmatrix} =
  \begin{pmatrix}
    0 \\
    - \gamma_t v
  \end{pmatrix}
\end{equation*}

\subsection{Step 2: $\nabla \cdot \Sigma (z)$}

Assuming that $\gamma_t$ and $\sigma_t$ are constant with respect to $v$, i.e.\ that $\nabla_v (\gamma_t \sigma_t^2) = 0$, we have
\begin{equation*}
  \nabla \cdot \Sigma (z) =
  \begin{pmatrix}
    \nabla_x (0) + \nabla_v (0) \\
    \nabla_x (0) + \nabla_v (\gamma_t \sigma_t^2) \\
  \end{pmatrix} =
  \begin{pmatrix}
    0 \\
    0
  \end{pmatrix}
\end{equation*}

\subsection{Step 3: Choice of $\xi(z)$}

Similarly to \cite{duncan2017}, we propose using the function
\begin{equation*}
  \xi(z) =
  \begin{pmatrix}
    v \\
    \sigma_t^2 \nabla_x \log \pi(x)
  \end{pmatrix}
\end{equation*}

Next we check that $\nabla \cdot (\rho \xi) (z) = 0$. We have
\begin{equation*}
  \rho (z) \xi (z) =
  \begin{pmatrix}
    K v \pi (x) \exp \left( -\frac{v^2}{2\sigma_t^2} \right) \\
    K \sigma_t^2 [\nabla_x \log \pi(x)] \pi(x)  \exp\left( -\frac{v^2}{2\sigma_t^2} \right) \\
  \end{pmatrix}
\end{equation*}

Then, assuming once again that $\sigma_t$ does not depend on $x$ and $v$,
\begin{equation*}
  \nabla \cdot (\rho \xi) (z) = K v \exp \left( -\frac{v^2}{2\sigma_t^2}\right) \nabla_x \pi(x) + K \sigma_t^2 [\nabla_x \log \pi (x)] \pi(x) \nabla_v \left[ \exp \left( -\frac{v^2}{2\sigma_t^2}\right) \right].
\end{equation*}

Substituting
\begin{equation*}
  \nabla_x \log \pi (x) = \frac{\nabla_x \pi(x)}{\pi(x)}
\end{equation*}
and
\begin{equation*}
  \nabla_v \left[ \exp \left( -\frac{v^2}{2\sigma_t^2}\right) \right] = -\frac{v}{\sigma_t^2} \exp \left( -\frac{v^2}{2\sigma_t^2}\right)
\end{equation*}
we find
\begin{equation*}
  \nabla \cdot (\rho \xi) (z) = 0
\end{equation*}
as required.

\subsection{Step 4: check condition}

Finally, we use the choice of $\xi$ proposed above to check the sufficient condition for stationarity given by \cite{duncan2017} and shown in Equation \ref{eqn:duncan_cond}. We have
\begin{align*}
  \Sigma(z) \nabla \log \rho (z) + \nabla \cdot \Sigma (z) + \xi(z)
  & =
    \begin{pmatrix}
      0 \\
      - \gamma_t v
    \end{pmatrix} +
    \begin{pmatrix}
      0 \\
      0
    \end{pmatrix} +
    \begin{pmatrix}
      v \\
      \sigma_t^2 \nabla_x \log \pi(x)
    \end{pmatrix} \\
  & =
    \begin{pmatrix}
      v \\
      - \gamma_t v + \sigma_t^2 \nabla_x \log \pi (x)
    \end{pmatrix} \\
  & = a(z)
\end{align*}
as required.

\subsection{Summary}

We have shown that the distribution $\rho$ is stationary for the underdamped Langevin model that we consider, under the condition that $\nabla_x \sigma_t = \nabla_v \sigma_t = \nabla_v \gamma_t = 0$. That is, the speed parameter $\sigma_t$ should not depend on the current location or velocity, and the friction parameter $\gamma_t$ should not depend on the current velocity.

We have focused on the one-dimensional case, but the proof generalises immediately to the $d$-dimensional isotropic case, with the stationary distribution
\begin{equation*}
    \rho(z) = K \pi(x) \exp \left( - \frac{\lVert v \rVert^2}{2\sigma_t^2}\right).
\end{equation*}

\section{Derivation of transition density}
\label{app:transition}

This appendix follows the calculations of \cite{cheng2018} to find the transition density of the discretised underdamped Langevin process, which we use to approximate the dynamics of the true process over short time intervals. Let $(X_t, V_t)^\intercal$ be a discretised underdamped Langevin process, defined as the solution to
\begin{equation*}
  \begin{cases}
    d X_t = V_t\ dt \\
    dV_t = -\gamma V_t\ dt + \sigma^2 \nabla \log \pi (X_0)\ dt + \sqrt{2 \gamma} \sigma\ dW_t
  \end{cases}
\end{equation*}
for small $t > 0$. The only difference with the equation for the exact process is that we use $\nabla \log \pi (X_0)$ in the drift term---i.e., the gradient is assumed constant between 0 and $t$.

\subsection{Solution}

\subsubsection{Velocity process}

We first notice that the second SDE can be written
\begin{align*}
  & dV_t = -\gamma V_t\ dt + \sigma^2 \nabla \log \pi (X_0)\ dt + \sqrt{2 \gamma} \sigma\ dW_t \\
  \Rightarrow\quad & dV_t + \gamma V_t dt =  \sigma^2 \nabla \log \pi (X_0)\ dt + \sqrt{2 \gamma} \sigma\ dW_t 
\end{align*}
and that the left-hand side is also
\begin{align*}
  dV_t + \gamma V_t dt & = e^{-\gamma t} (e^{\gamma t}\ dV_t + e^{\gamma t} \gamma V_t\ dt)\\
                       & = e^{-\gamma t} d(e^{\gamma t} V_t)
\end{align*}

Combining these two results,
\begin{align*}
  d(e^{\gamma t} V_t) = e^{\gamma t} \left( \sigma^2 \nabla \log \pi(X_0)\ dt + \sqrt{2\gamma} \sigma\ dW_t \right)
\end{align*}

We integrate both sides between 0 and $t$,
\begin{align*}
  & e^{\gamma t} V_t - V_0 = \int_{s = 0}^t e^{\gamma s} \left( \sigma^2 \nabla \log \pi(X_0)\ ds + \sqrt{2\gamma} \sigma\ dW_s \right) \\
  \Rightarrow\quad & V_t = V_0 e^{-\gamma t} + \sigma^2 \nabla \log \pi(X_0) \int_{s=0}^t e^{-\gamma (t-s)}\ ds + \sqrt{2\gamma} \sigma \int_{s=0}^t e^{-\gamma(t-s)}\ dW_s
\end{align*}
where
\begin{equation*}
  \int_{s=0}^t e^{-\gamma(t-s)}\ ds = \frac{1 - e^{-\gamma t}}{\gamma}
\end{equation*}

Finally:
\begin{equation*}
  V_t = e^{-\gamma t} V_0 + (1 - e^{-\gamma t}) \frac{\sigma^2 \nabla \log \pi (X_0)}{\gamma} + \sqrt{2\gamma} \sigma \int_{s=0}^t e^{-\gamma (t - s)}\ dW_s
\end{equation*}

\subsubsection{Location process}

The position is the integral of velocity, i.e.,
\begin{align*}
  X_t & = X_0 + \int_{u=0}^t V_u\ du \\
      & = X_0 + V_0 \int_{u=0}^t e^{-\gamma u}\ du + \frac{\sigma^2 \nabla \log \pi (X_0)}{\gamma} \int_{u=0}^t (1 - e^{-\gamma u})\ du + \sqrt{2\gamma} \sigma \int_{u=0}^t \int_{s=0}^u e^{-\gamma(u - s)}\ dW_s\ du
\end{align*}
where
\begin{equation*}
  \int_{u=0}^t e^{-\gamma u}\ du = \frac{1 - e^{-\gamma t}}{\gamma},
\end{equation*}
\begin{equation*}
  \int_{u=0}^t (1 - e^{-\gamma u})\ du = t - \frac{1 - e^{-\gamma t}}{\gamma},
\end{equation*}
and
\begin{align*}
  \int_{u=0}^t \int_{s=0}^u e^{-\gamma(u - s)}\ dW_s\ du & = \int_{s=0}^t \int_{u=s}^t  e^{-\gamma(u - s)}\ du\ dW_s \\
   & = \int_{s=0}^t \frac{1 - e^{-\gamma (t-s)}}{\gamma}\ dW_s
\end{align*}

Finally,
\begin{equation*}
  X_t = X_0 + (1 - e^{-\gamma t} ) \frac{V_0}{\gamma} + \frac{\sigma^2 \nabla \log \pi(X_0)}{\gamma} \left[ t - \frac{1 - e^{-\gamma t}}{\gamma} \right] + \frac{\sqrt{2} \sigma}{\sqrt{\gamma}} \int_{s = 0}^t (1 - e^{-\gamma (t - s)})\ dW_s
\end{equation*}

\subsection{Expectations}

As the expectation of the Ito integral is zero, we have
\begin{equation*}
  E[V_t] =  e^{-\gamma t} V_0 + (1 - e^{-\gamma t}) \frac{\sigma^2 \nabla \log \pi (X_0)}{\gamma}
\end{equation*}
and
\begin{equation*}
  E[X_t] = X_0 + (1 - e^{-\gamma t} ) \frac{V_0}{\gamma} + \frac{\sigma^2 \nabla \log \pi(X_0)}{\gamma} \left[ t - \frac{1 - e^{-\gamma t}}{\gamma} \right]
\end{equation*}

\subsection{Variances}

The variance of the velocity process is
\begin{align*}
  Var[V_t] & = E \left[ (V_t - E[V_t] )^2 \right] \\
           & = 2\gamma \sigma^2 E \left[ \left( \int_{s=0}^t e^{-\gamma (t - s)}\ dW_s \right)^2 \right] \\
           & = 2 \gamma \sigma^2 \int_{s=0}^t e^{-2\gamma (t - s)}\ ds \\
           & = 2\gamma \sigma^2 \times \frac{1 - e^{-2\gamma t}}{2\gamma} \\
           & = \sigma^2 (1 - e^{-2\gamma t})
\end{align*}
where the third equality follows from the Ito isometry.

For the location process, we have
\begin{align*}
  Var[X_t] & = E \left[ (X_t - E[X_t])^2 \right] \\
           & = \frac{2 \sigma^2}{\gamma} E \left[ \left( \int_{s=0}^t (1 - e^{-\gamma(t-s)}\ dW_s \right)^2 \right] \\
           & = \frac{2 \sigma^2}{\gamma} \int_{s = 0}^t (1 - e^{-\gamma(t-s)})^2\ ds \\
           & = \frac{2 \sigma^2}{\gamma} \times \frac{1}{2\gamma} (4 e^{-\gamma t} - e^{-2\gamma t} + 2 t \gamma - 3) \\
           & = \sigma^2 \left[ \frac{2t}{\gamma} - \frac{e^{-2\gamma t}}{\gamma^2} - \frac{3}{\gamma^2} + \frac{4 e^{-\gamma t}}{\gamma^2} \right]
\end{align*}

Similarly, the covariance is
\begin{align*}
  Cov [X_t, V_t] & = E[(X_t - E[X_t]) (V_t - E[V_t])] \\
                 & = E \left[ \left( \sqrt{2\gamma} \sigma \int_{s=0}^ t e^{-\gamma(t - s)}\ dW_s \right) \left( \frac{\sqrt{2} \sigma}{\sqrt{\gamma}} \int_{s=0}^t (1 - e^{-\gamma (t-s)})\ dW_s \right) \right] \\
                 & = 2 \sigma^2 \int_{s=0}^t e^{-\gamma (t-s)} (1 - e^{-\gamma (t-s)})\ ds \\
                 & = 2\sigma^2 \times \frac{1}{2\gamma} (1 - 2 e^{-\gamma t} + e^{-2\gamma t}) \\
                 & = \frac{\sigma^2}{\gamma} (1 - 2 e^{-\gamma t} + e^{-2\gamma t})
\end{align*}

\subsection{Summary}

By property of the Ito integral, $X_t$ and $V_t$ are Gaussian, and their joint distribution is fully described by the means and (co)variances found above. We focused on the 1-dimensional case above for simplicity, but the result directly applies to the $d$-dimensional isotropic case, where the dimensions are described by independent normal transition densities.

If we denote as $Z_t = (X_{t1}, V_{t1}, \dots, X_{td}, V_{td})^\intercal$ the joint position and velocity process, $I_d$ the $d \times d$ identity matrix, $\otimes$ the Kronecker product, and $\nabla_j$ the $j$-th component of the gradient with respect to location, then the transition density of the $d$-dimensional isotropic process is
\begin{equation*}
  Z_t \mid \{ X_0 = x_0, V_0 = v_0 \} \sim N(\mu,\ I_d \otimes Q)
\end{equation*}
where $\mu \in \mathbb{R}^{2d}$ with, for $j \in \{ 1, \dots, d\}$,
\begin{equation*}
    \begin{cases}
        \mu_{2(j-1)+1} = x_{0j} + \frac{1 - e^{-\gamma t}}{\gamma} v_{0j} + \frac{\sigma^2}{\gamma} \left[ t - \frac{1 - e^{-\gamma t}}{\gamma} \right] \nabla_j \log \pi(x_0)  \\
        \mu_{2j} = e^{-\gamma t} v_{0j} + \frac{\sigma^2}{\gamma} (1 - e^{-\gamma t}) \nabla_j \log \pi (x_0)
    \end{cases}
\end{equation*}
and
\begin{equation*}
  Q =
  \begin{pmatrix}
    \sigma^2 \left[ \frac{2t}{\gamma} - \frac{e^{-2\gamma t}}{\gamma^2} - \frac{3}{\gamma^2} + \frac{4 e^{-\gamma t}}{\gamma^2} \right] & \frac{\sigma^2}{\gamma} (1 - 2 e^{-\gamma t} + e^{-2\gamma t}) \\
     \frac{\sigma^2}{\gamma} (1 - 2 e^{-\gamma t} + e^{-2\gamma t}) & \sigma^2 (1 - e^{-2\gamma t})
  \end{pmatrix}
\end{equation*}

\section{Overdamped Langevin process as a limiting case}
\label{app:overdamped}

We first define $\tilde\sigma = \sigma \sqrt{\varepsilon}$ and $\tilde\gamma = \gamma \varepsilon$ for some $\epsilon > 0$, and we consider the situation where $\varepsilon \rightarrow 0$ (i.e., $\sigma \rightarrow \infty$, and $\sigma^2/\gamma$ is fixed). The SDE for $V_t$ (Equation \ref{eqn:model}) can be rewritten
\begin{equation*}
  dV_t = - \frac{\tilde\gamma}{\varepsilon} V_t\ dt + \frac{\tilde\sigma^2}{\varepsilon} \nabla \log[\pi(X_t)]\ dt + \frac{\sqrt{2\tilde\gamma} \tilde\sigma}{\varepsilon}\ dW_t
\end{equation*}
and, if we multiply both sides by $\varepsilon$,
\begin{equation*}
  \varepsilon dV_t = - \tilde\gamma V_t\ dt + \tilde\sigma^2 \nabla \log[\pi(X_t)]\ dt + \sqrt{2\tilde\gamma} \tilde\sigma\ dW_t
\end{equation*}

In the limit $\varepsilon \rightarrow 0$, the left-hand side tends to zero, and this equation becomes
\begin{equation*}
  \tilde\gamma V_t\ dt = \tilde\sigma^2 \nabla \log[\pi(X_t)]\ dt + \sqrt{2\tilde\gamma} \tilde\sigma\ dW_t
\end{equation*}

We plug this into the equation $dX_t = V_t dt$ (Equation \ref{eqn:model}), and we get
\begin{equation*}
  dX_t = \frac{\tilde\sigma^2}{\tilde\gamma} \nabla \log [\pi(X_t)]\ dt + \frac{\sqrt{2} \tilde\sigma}{\sqrt{\tilde\gamma}}\ dW_t.
\end{equation*}

This equation defines an overdamped (first-order) Langevin diffusion process, where the parameter $\zeta = \tilde\sigma/\sqrt{\tilde\gamma} = \sigma/\sqrt{\gamma}$ measures the diffusion speed \citep{michelot2019langevin}.

\section{Link to step selection functions}
\label{app:ssf}

Step selection functions are popular discrete-time models of animal movement and habitat selection, where the distribution of the position $x_{i+1}$ at time $t_{i+1}$ given previous positions $x_{1:i}$ is typically in the form
$$
f(x_{i+1} \mid x_{1:i} ) \propto \exp \left[ \sum_{k=1}^K \eta_k h_k(x_{i+1}, x_{1:i}) \right]
$$
where the $h_k$ are environmental or movement variables (e.g., spatial covariate, or length of step from $x_i$ to $x_{i+1}$). 

In this appendix, we derive an approximation of the discrete-time transition kernel of the underdamped Langevin model for small time intervals, and write it in the form of a step selection function. For simplicity, we denote as $X_1, X_2, \dots$ and $V_1, V_2, \dots$ the values of the processes $(X_t)$ and $(V_t)$, respectively, at regular observation times $t_1 < t_2 < \dots$. We denote as $\Delta = t_{i+1} - t_i$ the constant time interval, and assume that $\Delta$ is small.

Based on the results from Appendix \ref{app:transition}, the approximate transition density of the velocity process is
\begin{equation*}
    V_{i+1} \mid v_i, x_i \sim N \left( e^{-\gamma\Delta} v_i + (1-e^{-\gamma\Delta})\frac{\sigma^2}{\gamma} \nabla \log \pi(x_i),\ \sigma^2 (1-e^{-2\gamma\Delta}) I_d \right),
\end{equation*}
where $I_d$ is the $d \times d$ identity matrix. We combine this with a first-order discretisation of the velocity $V_i = (X_i - X_{i-1})/\Delta$, to find
\begin{equation*}
    X_{i+1} \mid x_i, x_{i-1} \sim N \left( x_i + e^{-\gamma\Delta}(x_i - x_{i-1}) + \Delta(1-e^{-\gamma\Delta}) \frac{\sigma^2}{\gamma} \nabla \log \pi(x_i),\ \sigma^2 \Delta^2 (1-e^{-2\gamma\Delta}) I_d \right)
\end{equation*}

The approximate transition density of the position process is therefore of the form
\begin{equation}
\label{eqn:approx_pdf}
f(x_{i+1} \mid x_i, x_{i-1}) \propto \exp \left[ -\frac{a^\intercal a}{2\sigma^2\Delta^2(1-e^{-2\gamma\Delta})} \right]
\end{equation}
where
\begin{equation*}
    a = (x_{i+1} - x_i) - e^{-\gamma\Delta} (x_i - x_{i-1}) - \Delta (1-e^{-\gamma\Delta}) \frac{\sigma^2}{\gamma} \nabla \log \pi(x_i).
\end{equation*}
and where the constant of proportionality does not depend on $x_{i+1}$. Expanding, and grouping together the terms that do not depend on $x_{i+1}$, we can write
\begin{equation*}
    a^\intercal a = \lVert x_{i+1} - x_i \rVert^2 - 2 e^{-\gamma\Delta} (x_{i+1} - x_i)^\intercal (x_i - x_{i-1}) - 2\Delta (1-e^{-\gamma\Delta}) \frac{\sigma^2}{\gamma}(x_{i+1} - x_{i})^\intercal \nabla \log \pi(x_i).
\end{equation*}

We denote as $L_i = \lVert x_{i+1} - x_i \rVert$ the step length between the two positions $x_i$ and $x_{i+1}$, and we denote as $\varphi_i$ the turning angle (i.e., change in direction) between the vectors $x_i - x_{i-1}$ and $x_{i+1} - x_i$. By property of the inner product, we have $(x_{i+1} - x_i)^\intercal (x_i - x_{i-1}) = L_{i-1} L_i \cos(\varphi_i)$.

Using a linear approximation of $\log \pi$ in a neighbourhood of $x_i$ and $x_{i+1}$, we have
$$
\log [\pi(x_{i+1})] \approx \log[\pi(x_i)] + (x_{i+1} - x_i)^\intercal \nabla \log[\pi(x_i)]
$$
such that $(x_{i+1} - x_i)^\intercal \nabla \log[\pi(x_i)] \approx \log[\pi(x_{i+1})] - \log[\pi(x_{i})]$.

Combining all these expressions we find
\begin{equation*}
    a^\intercal a = L_i^2 - 2e^{-\gamma\Delta} L_{i-1} L_i \cos(\varphi_i) - 2\Delta(1-e^{-\gamma\Delta}) \frac{\sigma^2}{\gamma} \log \pi(x_{i+1}) + C'
\end{equation*}
where $C'$ does not depend on $x_{i+1}$. Substituting this expression in Equation \ref{eqn:approx_pdf} gives the approximation
\begin{align*}
    f(x_{i+1} \mid x_i, x_{i-1}) \propto \exp \Bigg[ 
        \frac{-1}{2\Delta^2\sigma^2(1-e^{-2\gamma\Delta})} L_i^2 + 
        \frac{e^{-\gamma\Delta}}{\Delta^2\sigma^2(1-e^{-2\gamma\Delta})} L_{i-1} L_i \cos(\varphi_i) + \\ 
        \frac{1 - e^{-\gamma\Delta}}{\Delta \gamma (1-e^{-2\gamma\Delta})} \log \pi(x_{i+1}) \Bigg].
\end{align*}

We can write this approximate discrete-time transition density in terms of spatial covariates using $\pi(x) \propto \exp(\beta_1 \psi_1(x) +\dots \beta_J \psi_J(x))$ as proposed in Equation \ref{eqn:RSF}, and we get
\begin{align}
    f(x_{i+1} \mid x_i, x_{i-1}) \propto \exp \Bigg[ 
        \frac{-1}{2\Delta^2\sigma^2(1-e^{-2\gamma\Delta})} L_i^2 + 
        \frac{e^{-\gamma\Delta}}{\Delta^2\sigma^2(1-e^{-2\gamma\Delta})} L_{i-1} L_i \cos(\varphi_i) + \nonumber \\ 
        \sum_{k=1}^K \frac{(1 - e^{-\gamma\Delta}) \beta_k}{\Delta \gamma (1-e^{-2\gamma\Delta})} \psi_k(x_{i+1}) \Bigg].
        \label{eqn:ssf}
\end{align}

Equation \ref{eqn:ssf} can be viewed as a step selection function that includes (a) the squared step length with a negative coefficient, representing avoidance of long steps due to energetic constraints, (b) the product $L_{i-1}L_i\cos(\varphi_i)$ with a positive coefficient, which captures movement persistence, and (c) the spatial covariates $\psi_1, \dots, \psi_K$, for selection or avoidance of habitat features.

For $\Delta \to 0$, a first-order Taylor expansion of the coefficients gives
\begin{align*}
    \frac{-1}{2\Delta^2\sigma^2(1-e^{-2\gamma\Delta})} & \approx \frac{-1}{4\gamma\sigma^2\Delta^3} \\
    \frac{e^{-\gamma\Delta}}{\Delta^2\sigma^2(1-e^{-2\gamma\Delta})} & \approx \frac{1}{2\gamma\sigma^2\Delta^3} \\
    \frac{(1 - e^{-\gamma\Delta}) \beta_k}{\Delta \gamma (1-e^{-2\gamma\Delta})} & \approx \frac{\beta_k}{2\gamma\Delta}
\end{align*}

As expected, the negative coefficient for $L_i^2$ decreases (in absolute value) as the speed parameter $\sigma$ increases, because a fast-moving animal is less reluctant to take long steps, and the positive coefficient for $L_{i-1} L_i \cos(\varphi_i)$ decreases as the friction parameter $\gamma$ increases, because higher friction leads to lower movement persistence. 

Interestingly, the habitat selection coefficients are predicted to decrease as the time interval $\Delta$ increases. This contrasts with previous theoretical work on step selection parameters in the absence of directional persistence, where the estimated selection coefficients were found to increase from $\beta_k/2$ at fine observation scales to $\beta_k$ at long observation scales \citep{moorcroft2008, potts2023}. This suggests that directional persistence is an important factor to understand how spatial distributions arise from small-scale movements. Our results coincide with those previous findings as $\Delta \to 0$ only in the case where $\gamma = 1/\Delta$, which implies $\gamma \to \infty$, corresponding to infinite friction and no movement persistence.

\end{document}